# Extending Scatterplots to Scalar Fields

Shenghui Cheng[1], Pengcheng Cui[2] and Klaus Mueller[1]

[1]Visual Analytics and Imaging Lab, Computer Science Department, Stony Brook University  [2]Mathematics Department, Shandong University

**ABSTRACT**

Embedding high-dimensional data into a 2D canvas is a popular strategy for their visualization. It allows users to appreciate similarity relationships among the data points by their spatial organization on the 2D display. In this work we consider the case where the collection of data points also serves as a scalar field for one of the dataset's attributes, essentially forming samples of the attribute's continuous function in this embedded space. We study methods by which this continuous function can be estimated from the discrete samples, making certain assumptions on the function's smoothness in high-dimensional space. Our method allows users to create distance fields, iso-contours, topographic maps, and even extrapolations to embed the possibly odd-shaped point assembly into a filled rectangular region.

## 1 INTRODUCTION

Our paper in some sense bridges the boundary between scientific visualization (SciVis) and information visualization (InfoVis). In SciVis the data typically come on regular or semi-regular grids which tie the data to the underlying phenomena's spatial organization and continuity in the attributes measured in this space. This is different from InfoVis where the data are predominantly non-spatial and often live in a high-dimensional attribute space that is embedded in the two-dimensional canvas, giving rise to irregular point clouds in the form of scatterplots.

In SciVis, color and brightness typically encode the value of the primary attribute, such as temperature, density, speed, etc. In InfoVis, on the other hand, color and brightness are predominantly employed to encode the membership of points in certain clusters. Additional variables are most often encoded as point size – a popular example being Gapminder which codes the magnitude of a certain entity as size in its characteristic animated display. Using size to encode an attribute's value, however, limits the resolution of the display.

The use of color and brightness to encode a primary attribute of the data – as opposed to the aforementioned cluster membership – is a frequent practice in SciVis. These types of displays are often referred to as *scalar fields*. Scalar fields are defined over a continuous domain and typically have a smooth and continuous appearance. An example is the variation of pressure or temperature over a geometric shape such as an airplane wing. In this work, we aim to adapt the notion of scalar field to InfoVis displays – namely to encode the value of a chosen attribute or interest. This brings the advantage that in contrast to using node size for this purpose, employing color or brightness does not limit the display's resolution.

However, a significant obstacle in this endeavor is the inherent spatial disorganization of the InfoVis non-spatial data. Overcoming this limitation is at the heart of our paper. To achieve our goal, we propose a regularizing non-linear transform of the spatial organization of the data. This transform creates a smooth transition in the color-coded variable making it easy to see trends in the context of the other variables. Similar to scalar fields the visualizations we create are dense and not scattered. This enables other useful types of visualizations, such as iso-contours, topographic maps, and even extrapolations.

## 2 APPROACH

To illustrate the problem, we will use the 7-D UCI Auto MPG dataset. It consists of 392 cars with 7 attributes – miles per gallon (MPG), #cylinders (CYL), horsepower (Hpower), weight, acceleration (Accel), year, and origin. Coloring a scatterplot of cars over two attributes with a third attribute, say horsepower, will give rise to a random confetti-like arrangement of colored points. The same is true when attempting to color a 2D layout generated by multi-dimensional scaling (MDS). Fig. 1 shows such a display where we mapped HPower to brightness. We can easily observe that bright and less bright points are distributed across the display without clear structure. This makes it difficult to derive insight from this plot, especially when it comes to HPower.

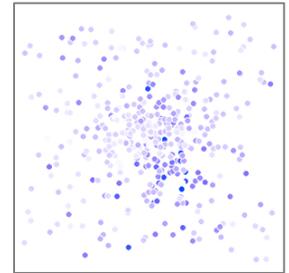

Figure 1. An MDS layout of the car data set with HPower mapped to brightness.

We note that this phenomenon is less likely to occur in choropleth maps, which are also often used in InfoVis. Choropleth maps, however, are different from multivariate scatterplots as the data used there have a geo-spatial component and hence color-coded variables tend to have a smooth spatial distribution.

Essentially the problem arises because the MDS-layout of the data points (blue in Fig. 1) is only mildly associated with the value of the attribute HPower. Rather, it has been derived by preserving the pairwise distances of the data points in high-dimensional space spanned by all attributes, minimizing the MDS stress function. To derive a better and more organized display we take advantage of a framework we developed recently [1], called the *data context map*. The data context map creates a composite layout by augmenting the original data distance matrix used in MDS by three additional matrices – the attribute correlation matrix and two data-attribute affinity matrices where entries map to the degree of affinity a data point has with respect to certain attribute. MDS optimization is then used to embed this composite matrix into 2D canvas space where three similarities are consistently preserved – data, attribute, and data/attribute.

### 2.1 Creating a Continuous Map via AKDE Regression

Using the layout of the composite matrix organizes the data points in terms of the target attribute, here HPower, but we still only have a set of data points placed at irregular canvas locations, while a scalar field is a continuous function. In order to fill the empty space we require an interpolation method – and one that can deal with non-regular point distributions. Quite a few such interpolation techniques exist, such as nearest neighbor, linear, natural neighbor, etc. We found that but none of these fully satisfied the following two important criteria: (1) the values in the estimation area should distribute continuously and smoothly, and (2) the original data point values should be maintained.

---
* {shecheng, mueller}@cs.stonybrook.edu

Hence, for this purpose, we devised a new estimation method called OIE-AKDE, which stands for an original (O), interpolating (I), and extrapolating (E) variant of adaptive kernel density estimation regression (AKDE) [2]. It first estimates the density of samples and computes the bandwidth of each sample with AKDE. Then it "calculates" different value distributions for each sample. Computing the impact or diffusion factors of each sample in each dimension, we gain an expression for the value distribution in the given estimation area.

A helpful illustration often used in scalar fields is the *iso-contour* – a closed curve that indicates a certain level of the scalar attribute. A set of iso-contours, one for each level at some level spacing, can then function as a topographic map. This map enables and easy assessment of the scalar attribute's level at a certain spatial location. It also enables users to visually spot areas where the scalar variable changes rapidly (areas with dense contours) or only varies gradually (areas with sparse contours).

Fig. 2 shows a visualization of the UCI Auto MPG dataset using the method just described. From this plot, we can observe that the cars near the attribute "HPower" have high values for that attribute and vice versa. While we can also make similar observations for other attributes by comparing the data items with the placement of their respective nodes, the contour lines computed for the "Hpower" attribute reveal additional information. For example, if a user desires a car with "HPower" between 120~140, he or she can simply find the car of interest based on the iso-contour, e.g., a red circled one - "Peugeot 604s" with its values displaying on the right block.

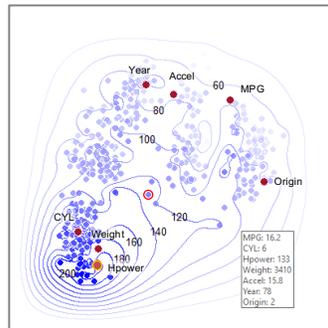

Figure 2. An MDS layout using the composite data matrix for MDS based layout and showing equal HPower values as iso-contours. Other car attributes are also shown as embedded nodes.

The visualization shown in Fig. 2 is a hybrid of the traditional contour map (using the "HPower" attribute as the scalar variable), an MDS-optimized scatterplot display for the data items (blue points), and a correlation map for all the attributes (red points). The MDS-optimized scatterplot display links the scalar field to its new setting in the InfoVis domain where the goal is to visualize data acquired at irregular locations (as opposed to functions defined in a continuous space). Conversely, the attribute correlation display links the scalar field to the high-dimensional data domain.

## 3 CASE STUDY – SELECTING A UNIVERSITY OF BEST FIT

Selecting a good university is significant for perspective students. The university typically consists of multiple attributes. How to balance these attributes is difficult for students. We obtained a dataset containing 46 universities along with 14 attributes of interest: academics, athletics, housing, location, nightlife, safety, transportation, weather, score, tuition, dining, PhD/faculty, population, and income. For the perspective student Peter, he considers the following criteria: low tuition, good academic and good athletic. He would like to choose a university with good academics (>9), good athletics (>9) and low tuition (<$18,000).

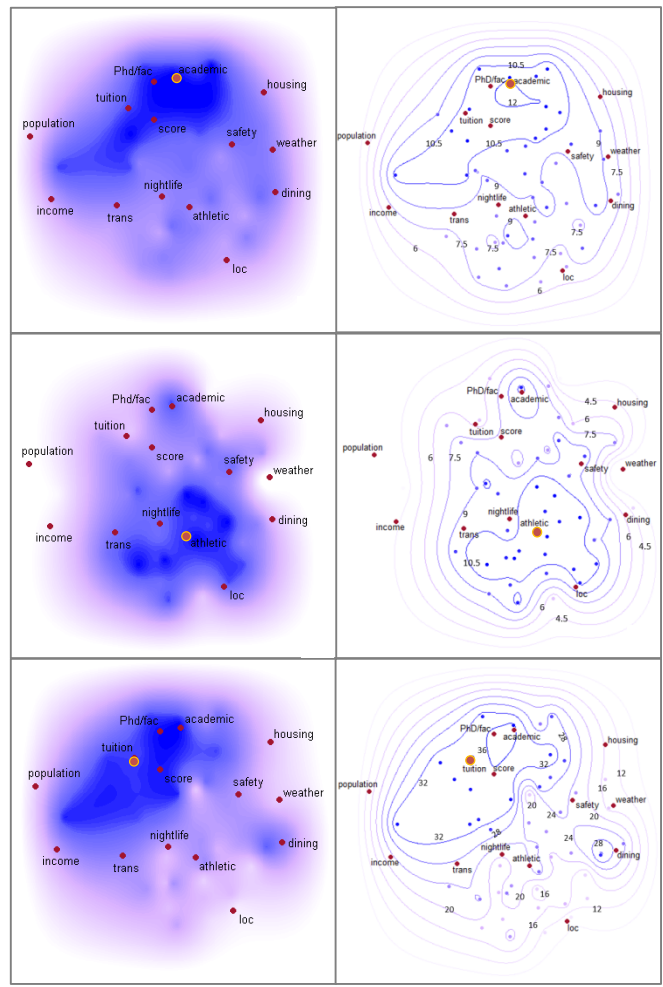

We now use our method to visualize the data set (see Fig. 3). The large red nodes represent the attributes while the small blue points represent the universities. We apply the OIE-AKDE method with to estimate the whole area on the attributes "academic", "athletic" and "tuition" respectively – see Fig.3a, c, e. We observe that the scalar field is smooth for all attributes. In order to force extrapolations, we consider the data values to grow smaller when we come closer to the border, and assume the values of all border points to be zero in each dimension. By means of this setting, the estimation function will bend to zero and fade to border. Fig. 3b, d, shows the associated contour maps.

**ACKNOWLEDGMENTS**

This research was partially supported by NSF grant IIS 1527200 and the MSIP (Ministry of Science, ICT and Future Planning), Korea, under the "IT Consilience Creative Program (ITCCP)" (NIPA-2013-H0203-13-1001) supervised by NIPA.

**REFERENCES**

[1] S. Cheng, K. Mueller, "The Data Context Map: Fusing Data and Attributes into a Unified Display," *IEEE Trans. on Visualization and Computer Graphics* 22(1): 121-130, 2016.
[2] P. Kerm, "Adaptive Kernel Density Estimation," *The Stata Journal*, 2:148–156, 2002.